\documentclass[prd,superscriptaddress,floatfix,amsfonts,amssymb,amsmath,showpacs,twocolumn]{revtex4}
\usepackage{bm}
\usepackage{amsfonts}
\usepackage{latexsym}
\usepackage[latin1]{inputenc}
\usepackage{graphicx}
\usepackage{amsmath}
\usepackage{palatino}
\usepackage{mathpazo}
\usepackage{textcomp}
\usepackage{booktabs}
\usepackage{dcolumn}
\usepackage{hyperref}
\hypersetup{colorlinks,citecolor=blue}
\usepackage{amsmath}
\usepackage{xcolor}
\usepackage{mathtools}
\usepackage{multirow}
\usepackage{float}
\usepackage{orcidlink}
 
\def\jnl@style{\it}
\def\aaref@jnl#1{{\jnl@style#1}}

\def\aaref@jnl#1{{\jnl@style#1}}

\def\aj{\aaref@jnl{AJ}}                   % Astronomical Journal
\def\apj{\aaref@jnl{ApJ}}                 % Astrophysical Journal
\def\apjl{\aaref@jnl{ApJ}}                % Astrophysical Journal, Letters
\def\apjs{\aaref@jnl{ApJS}}               % Astrophysical Journal, Supplement
\def\apss{\aaref@jnl{Ap\&SS}}             % Astrophysics and Space Science
\def\aap{\aaref@jnl{A\&A}}                % Astronomy and Astrophysics
\def\aapr{\aaref@jnl{A\&A~Rev.}}          % Astronomy and Astrophysics Reviews
\def\aaps{\aaref@jnl{A\&AS}}              % Astronomy and Astrophysics, Supplement
\def\mnras{\aaref@jnl{Mon.~Not.~Roy.~Astron.~Soc.}}             % Monthly Notices of the RAS
\def\prd{\aaref@jnl{Phys.~Rev.~D}}        % Physical Review D
\def\prc{\aaref@jnl{Phys.~Rev.~C}}  % Physical Review C
\def\prl{\aaref@jnl{Phys.~Rev.~Lett.}}    % Physical Review Letters
\def\qjras{\aaref@jnl{QJRAS}}             % Quarterly Journal of the RAS
\def\skytel{\aaref@jnl{S\&T}}             % Sky and Telescope
\def\ssr{\aaref@jnl{Space~Sci.~Rev.}}     % Space Science Reviews
\def\zap{\aaref@jnl{ZAp}}                 % Zeitschrift fuer Astrophysik
\def\nat{\aaref@jnl{Nature}}              % Nature
\def\aplett{\aaref@jnl{Astrophys.~Lett.}} % Astrophysics Letters
\def\apspr{\aaref@jnl{Astrophys.~Space~Phys.~Res.}} % Astrophysics Space Physics Research
\def\physrep{\aaref@jnl{Phys.~Rep.}}      % Physics Reports
\def\physscr{\aaref@jnl{Phys.~Scr}}       % Physica Scripta
\def\commat{\aaref@jnl{Comm.~Math.~Phys.}}              % Communications in Mathematical Physics
\def\science{\aaref@jnl{Science}}               % Science
\def\cqg{\aaref@jnl{Classical Quant.~Grav.}}            % Classical and Quantum Gravity
\def\jpcs{\aaref@jnl{JPCS}}                                     % Journal of Physics Conference Series
\def\ijmpd{\aaref@jnl{Int.~J.~Mod.~Phys.~D}}                    % International Journal of Modern Physics D
\def\grg{\aaref@jnl{Gen.~Relat.~Gravit.}}               % General Relativity and Gravitation
\def\rpp{\aaref@jnl{Rep.~Prog.~Phys.}}          % Reports on Progress in Physics
\def\npa{\aaref@jnl{Nucl.~Phys.~A}}        % Nuclear Physics A
\def\lrr{\aaref@jnl{Living Rev.~Rel.}}                   % Living reviews in relativity
\def\jcap{\aaref@jnl{J.~Cosmology Astropart.~Phys.}}    % Journal of cosmology and astroparticle physics
\def\rmp{\aaref@jnl{Rev.~Mod.~Phys.}}   %Reviews of modern physics
\def\epjc{\aaref@jnl{Eur.~Phys.~J. ~C}}
\def\plb{\aaref@jnl{Phys.~Lett .~B}}

\allowdisplaybreaks[1]
\begin{document}
\color{red}

\title{A Complete Cosmological Scenario in Teleparallel Gravity}

\author{Sanjay Mandal\orcidlink{0000-0003-2570-2335}}
 \email{sanjaymandal960@gmail.com}
\affiliation{ Department of Mathematics, Birla Institute of Technology and Science-Pilani,\\ Hyderabad Campus,Hyderabad-500078, India}
\author{P.K. Sahoo\orcidlink{0000-0003-2130-8832}}
 \email{pksahoo@hyderabad.bits-pilani.ac.in}
\affiliation{ Department of Mathematics, Birla Institute of Technology and Science-Pilani,\\ Hyderabad Campus,Hyderabad-500078, India}

\begin{abstract}

Teleparallel gravity is a modified theory of gravity in which the Ricci scalar $R$ of the Lagrangian replaced by the general function of torsion scalar $T$ in action. With that, cosmology in teleparallel gravity becomes profoundly simplified because it is second-order theory. The article present a complete cosmological scenario in $f(T)$ gravity with $f(T)=T+\beta(-T)^{\alpha}$, where $\alpha,$ and $\beta$ are model parameters. We present the profiles of energy density, pressure, and equation of state (EoS) parameter. Next to this, we employ statefinder diagnostics to check deviation from the $\Lambda$CDM model as well as the nature of dark energy. Finally, we discuss the energy conditions to check the consistency of our model and observe that SEC violates in the present model supporting the acceleration of the Universe as per present observation.

\end{abstract}



\maketitle

\section{Introduction}\label{I}

Several observations confirm that the universe is accelerating in every second \cite{Riess/1998,Perlmutter/1999,deBernardis/2000,Colless/2001,Perlmutter/2003,Spergel/2003,Peiris/2003,Tegmark/2004,Cole/2005,Springel/2006,Astier/2006,Riess/2007,Spergel/2007,Ade/2014,Komatsu}. The main gradient which is responsible for the acceleration of the universe is unknown so-called `dark energy'. Nowadays, one of the difficult problems of modern cosmology and particle physics is to identify the properties of dark energy. The first time, it was described correctly by adding a cosmological constant ($\Lambda$) to Einstein's equation, but its magnitude is unwanted and unmotivated by the fundamental physics. Also, it has agreed that approximately $75\%$ of the total energy of the universe covered with dark energy. Furthermore, according to observations, it is represented by an EoS parameter $\omega\simeq -1$. Looking at all these things, cosmologists have been proposed a lot of proposals to overcome this problem \cite{Ratra/1998,Caldwell/1998,Buchert/2000,Picon/2001,Tomita/2001,Milton/2003,Hunt/2010,Easson/2011,Radicella/2012,Pavn/2013,Pandey/2017,Pandey/2019}.\newline
The modification of general relativity was a great idea to describe dark energy. As a result, several modified theories have proposed in the literature. Because it presents the nature of dark energy as a geometrical property of the universe, also, all the modified gravity theories are connected to the modification of Einstein-Hilbert action \cite{Paliathanasis/2016}. The modified theories of gravity such as $f(R)$ gravity, $f(R,T)$ gravity, Gauss-Bonnet gravity, etc. are widely used in modern cosmology (some interesting results reported in \cite{Nojiri/2017,Bhatti/2020,Santos/2019,Moraes/2016,Moraes/2019a,Moraes/2017,Elizalde/2019,Sahoo/2018,Elizalde/2018,Moraes/2018,Sharif/2019,Sahoo/2018a,Sahoo/2017,Azizi/2013,Alves/2016,Zaregonbadi/2016,Yousaf/2019}).\newline
A well established modified theory of gravity, which has attracted the interests of the cosmologists, is so-called $f(T)$ teleparallel gravity \cite{Bengochea/2009,Ferraro/2007,Linder/2010,Cai/2016}. Teleparallel gravity motivated by the generalization of $f(R)$ gravity in which the arbitrary function $f$ of Ricci scalar $R$ replaced by the arbitrary function of torsion scalar $T$ in action. We used the conventional torsionless Levi-Civita connection in general relativity, whereas the curvature-less Weitzenb$\ddot{o}$ck connection used in teleparallel gravity to describe the effects of gravitation in terms torsion instead of curvature \cite{Paliathanasis/2016,Hayashi/1979,Tsamparlis/1979,Arcos/2004}.\newline
The linear forms of $f(T)$ are the teleparallel equivalent of general relativity (TEGR) \cite{Einstein/1928}. However, the physical interpretation of $f(T)$ gravity is different from $f(R)$ gravity. Also, the Ricci scalar $R$ of $f(R)$ gravity contains the second-order derivatives of the metric tensor, whereas the torsion scalar $T$ of $f(T)$ gravity contains only the first-order derivatives of the vierbeins. Therefore, it is easy to find the exact solutions of the cosmological models in $f(T)$ gravity than the other modified theories of gravity. Although it is a simple modified theory, there are few exact solutions proposed in the literature. Some power-law solutions in Friedmann-Lema\^{i}tre-Robertson-Walker spacetime and anisotropic spacetime have found in \cite{Atazadeh/2012,Basilakos/2013,Rodrigues/2015}. Solution for static spherically symmetric spacetime and Bianchi I spacetime reported in \cite{Paliathanasis/2016,Paliathanasis/2014,Capozziello/2013}.\newline
In $f(T)$ gravity, the study of cosmological scenarios is easier in comparison to other modified theories of gravity. So, it has incorporated to study the cosmological scenarios such as big bounce \cite{Cai/2011,Haro/2015,Haro/2016,Hanafy/2017}, inflationary model \cite{Bamba/2016}, late time cosmic acceleration \cite{Bengochea/2009,Bengochea/2009a,Myrzakulov/2011}. Recently, spherical and cylindrical solution \cite{Capozziello/2020}, conformally symmetric traversable wormholes \cite{Singh/2020}, noether charge and black hole entropy \cite{Hammad/2019} in $f(T)$ gravity have been discussed. Moreover, assuming $f(T)$ as a power law function observational constraints \cite{Nunes/2016,Wu/2010}, dynamical system \cite{Yu/2010,Mirza/2017}, and structure formation \cite{Nunes/2018} have been studied in $f(T)$ gravity. \newline
The manuscript represented as follows: we started with a comparison study between GR anf teleparallel gravity in section \ref{a}. In section \ref{II}, we discuss the overview of $f(T)$ gravity and present the gravitational field equations in a spatially flat FLRW spacetime. In section \ref{III}, we discuss some kinematic variables such as Hubble parameter, deceleration parameter, which is followed by the construction of the cosmological model in $f(T)$ gravity in section \ref{IV}. Also, we present the expressions and behaviours for energy density, pressure, and EoS parameter. In section \ref{sd}, we discuss some geometrical diagnostics to distinguish our model from other dark energy models. The energy conditions for our model are discussed in section \ref{ecs}. And finally, in section \ref{f}, we present our results and conclusions. 

\section{General Relativity vs Teleparallel gravity}
\label{a}
The theory of Teleparallel gravity originated from Einstein's first work is inspired by the observations, which shows that a tetrad has 16 independent components. The metric tensor is described among these ten components, and thus describes gravity.
Also, the other six describe the electromagnetic  field \cite{Einstein/1928}. Later, it is found that the other six were components related to inertial observer/effects, rather than electromagnetism \cite{Cho/1976}. Here, we would like to note that in teleparallel gravity formulations, one should use the original concept of Einstein's teleparallelism; that is our modern way of describing a gravity where the spin connection vanishes. Therefore only the tetrad determines teleparallel gravity.
The key downside of this approach is that the anholonomy coefficients have replaced torsion tensors, which are not tensors under local Lorentz transformations, resulting in a violation of local Lorentz symmetry. However, this does not affect the field equations, and are still invariant under local Lorentz transformation \cite{Obukhov/2006}. This further confirms that the metric tensor obtained from this formalism is the right one \cite{Maluf/2013}. The critical difference is the existence of the teleparallel gravity tetrad field, which can be used to describe the Weitzenb$\ddot{o}$ck relation (in terms of torsion without curvature) and the Riemannian metric (in terms of Levi-Citiva connection). So, we can not say that curvature and tetrad are properties of the connection. Hence, multiple connections can be described in the same spacetime \cite{Cai/2016}. Besides, the gauge transformation plays a vital role in this formulation, as both Riemannian and teleparallel geometry on the same spacetime is generated in the presence of a non-trivial tetrad field. This universal property of gravitational interaction thus creates a connection between both gravitational structures and the gravitational interactions.

The same definition of gravity exists in teleparallel gravity, including torsion and curvature. We do have a common idea, though. In GR, curvature determines space-time geometry, which adequately explains the gravitational interaction. In contrast, teleparallel gravity uses torsion to describe the gravitational interaction, which is acted by force. The geodesic equations in the teleparallel equivalent to GR \cite{Andrade/1997} are comparable to the Lorentz force equation of electromagnetics. In addition, GR's teleparallel version preserves the geodesic structure. The gravitational interaction can thus be described in terms of curvature at GR or torsion in teleparallel gravity.\\
The formulation of any theory of gravity requires the following: one has to make some ad-hoc condition, insert experimental and/or observational information, follow the self-consistency of the Mathematical expressions. Nevertheless, in GR formulation, Einstein added the following conditions arise from the observation: metricity; Lorentz invariance; Equivalence Principle; causality. These conditions do not originate from the fundamental principle, but these contracted from the observation. In addition to this, Einstein imposed the symmetric, Levi-Citiva connection without any theoretical and observational background, and also considered torsion is zero and that geometry. Hence, gravity described by the curvature only. Finally, he ended up with an action having second order in metric and its derivatives. That action is called Einstein-Hilbert action with Ricci scalar R.\\
Furthermore, Einstein's gravity is accepted by the community, but it still faces a crucial question as to what kind of geometry to consider since there is much possibility to choose the correspondence. In the 20th century, Riemannian geometry was considered the most straightforward option to describe geometry. Rather than this, there are several possibilities to choose geometry. For instance, one could skip all the assumption made by Einstein such as metricity; Lorentz invariance; Equivalence Principle except the remaining conditions comes from the observational bound. However, one can safely describe the gravitational interaction through both curvature and torsion, this type of work done by Einstein as the Teleparallel Equivalent of GR (TEGR). In this case, the gravitational equations describe by the torsion are utterly the same as in GR. Therefore, GR and TEGR are equivalent; one can safely use one of them. When it comes to their advantage, TEGR has one advantage compared to GR, which was the successful calculation of the energy solution, whereas GR faces the problem like the use of pseudo-tensor \cite{Maluf/2002}. Nevertheless, other solutions are the same. We want to mention here that teleparallel gravity does not require the Equivalence Principle because the Weitzenb $\ddot{o}$ck connection describes its gravitational mechanism.\\
However, in the last two decades, researchers started to follow the modified gravity approach to study the accelerated phases of the universe, as it successfully describes the late time acceleration and the early time inflation. There are several proposed modified gravity theories. One can not refute another modified theory to present a new modified theory, because experiment/observation could only do that. Nonetheless, we must bear in mind that modified theory must be equivalent to the level of gravitational theory equations, and may not be equal to their modifications. Therefore, any possibility of alteration is worth investigating.
The modified curvature-based theories were 
the extension of the Einstein-Hilbert action. 
The Ricci scalar $R $replaced by the function 
generally depends on $R $combined with invariant curvature. As we discussed, this is not the only way to change the theories of gravity. One may start with the TEGR in particular, and then the Lagrangian torsion that is replaced by the function depends on $ T $ and the combination of different torsion quantities. The acceleration of $f(R)$ is the simplest based on curvature, while the simplest based on torsion is $f(T)$. The critical argument about these two modified theories is that given different gravitational mechanisms, GR and TEGR are identical at equation level.

\section{Teleparallel Gravity}\label{II}\label{II}

In this section, we briefly discuss the $f(T)$ gravity. The vierbein fields, $e_{\mu}(x^i)$ act as a dynamical variable for the teleparallel gravity. As usual, $x^i$ use to run over the space-time coordinates, and $\mu$ denotes the tangent space-time coordinates. At each point of the manifold, the vierbein fields form an orthonormal basis for the tangent space, which is presented by the line element of four-dimensional Minkowski space-time i.e., $e_{\mu}e_{\nu}=\eta_{\mu\nu}=$diag$(+1,-1,-1,-1)$. In vector component, the vierbein fields can be expressed as $e^i_{\mu}\partial_i$, and the metric tensor can be written as
\begin{equation}\label{1}
g_{\mu\nu}=\eta_{ij}e^i_{\mu}(x)e^j_{\nu}(x).
\end{equation}
Moreover, the vierbein basis follow the general relation $e^i_{\mu}e^{\mu}_j=\delta^i_j$ and $e^i_{\mu}e^{\nu}_i=\delta^{\nu}_{\mu}$.
In $f(T)$ gravity, the curvatureless Weitzenb$\ddot{o}$ck connection \cite{r/1923} defined as
\begin{equation}\label{2}
\hat{\Gamma}^{\gamma}_{\mu\nu}\equiv e^{\gamma}_i\partial_{\nu}e^i_{\mu}\equiv -e^i_{\mu}\partial_{\nu}e^{\gamma}_i.
\end{equation}
Using Weitzenb$\ddot{o}$ck connection one can write the non-zero torsion tensor as
\begin{equation}\label{3}
T^{\gamma}_{\mu \nu}\equiv \hat{\Gamma}^{\gamma}_{\mu\nu}-\hat{\Gamma}^{\gamma}_{\nu\mu} \equiv e^{\gamma}_i(\partial_{\mu} e^i_{\nu}-\partial_{\nu} e^i_{\mu}).
\end{equation}
The contracted form of the above torsion tensor can be written as \cite{Maluf/1994,Hayashi/1979,Arcos/2004}
\begin{equation}\label{4}
T\equiv S^{\mu \nu}_{\gamma}T^{\gamma}_{\mu \nu}\equiv \frac{1}{4}T^{\gamma \mu \nu}T_{\gamma \mu \nu}+\frac{1}{2}T^{\gamma \mu \nu}T_{\nu \mu \gamma}-T^{\gamma}_{\gamma \mu}T^{\nu \mu}_{\nu},
\end{equation}
where
\begin{equation}\label{5}
S^{\mu \nu}_{\gamma}=\frac{1}{2}(K^{\mu \nu}_{\gamma}+\delta^{\mu}_{\gamma}T^{\alpha \nu}_{\alpha}-\delta^{\nu}_{\gamma}T^{\alpha \mu}_{\alpha}) 
\end{equation}
be the superpotential tensor and the
difference between the Levi-Civita and Weitzenb$\ddot{o}$ck connections is the contortion tensor which is defined as
\begin{equation}\label{6}
K^{\mu \nu}_{\gamma}=-\frac{1}{2}(T^{\mu \nu}_{\gamma}-T^{\nu \mu}_{\gamma}-T^{\mu \nu}_{\gamma}).
\end{equation}
The extension of Einstein-Hilbert Lagrangian of $f(T)$ theory of gravity \cite{Cai/2016} (which is similar to $f(R)$ gravity extension from the Ricci scalar $R$ to $R+f(R)$ in the action), namely the teleparallel gravity term $T$ is replaced by $T+f(T)$, where $f(T)$ is an arbitrary function of $T$ as
\begin{equation}\label{7}
S=\frac{1}{16\pi G}\int[T+f(T)]e d^4x,
\end{equation}
where $e=$det$(e^i_\mu)=\sqrt{-g}$ and $G$ is the gravitational constant. Assume $k^2=8\pi G=M_p^{-1}$, where $M_p$ is the Planck mass.
By the variation of the total action $S+L_m$, here $L_m$ is the matter Lagrangian gives us the field equation for $f(T)$ gravity as
\begin{widetext}
\begin{equation}\label{8}
e^{-1}\partial_{\mu}(ee^{\gamma}_i S^{\mu \nu}_{\gamma})(1+f_T)-(1+f_T)e^{\lambda}_i T^{\gamma}_{\mu \lambda}S^{\nu \mu}_{\gamma}+e^{\gamma}_i S^{\mu \nu}_{\gamma}\partial_{\mu}(T)f_{TT}+\frac{1}{4}e^{\nu}_i[T+f(T)]=\frac{k^2}{2} e^{\gamma}_iT^{(M)\nu}_{\gamma},
\end{equation}
\end{widetext}
where $f_T=df(T)/dT$, $ f_{TT}=d^2f(T)/dT^2$, $T^{(M)\nu}_{\gamma}$ represents the energy-momentum tensor to the matter Lagrangian $L_m$.\\
Now we consider a flat FLRW universe with the metric as
\begin{equation}\label{9}
ds^2=dt^2-a^2(t)dx^{\mu} dx^{\nu},
\end{equation}
where $a(t)$ is the scale factor, which gives us
\begin{equation}\label{10}
e^{i}_{\mu}=diag(1,a,a,a).
\end{equation}
Moreover, assuming the energy-momentum tensor for the perfect fluid which takes the form
\begin{equation}\label{11}
T^{(M)}_{\mu\nu}=(\rho+p)u_{\mu}u_{\nu}-pg_{\mu\nu},
\end{equation}
where $\rho,p$ and $u^{\mu}$ be the energy density, pressure and the four velocity of the matter fluid, respectively.\\
Using equation \eqref{9} into the field equation \eqref{8}, we get the modified field equations as follows
\begin{equation}\label{12}
6H^{2}+f+Tf_T=16 \pi G\rho,
\end{equation}
\begin{equation}\label{13}
\dot{H}\left(1+f_{T}+2 T f_{T T}\right)=-4 \pi G(\rho+p) ,
\end{equation}
where $H\equiv \dot{a}/a$ be the Hubble parameter and `dot' represents the derivative with respect to $t$. Additionally, we have used the relation
\begin{equation}\label{14}
T=-6H^2,
\end{equation}
which holds for a FLRW Universe according to equation \eqref{4}.\\
%\begin{equation}
%\label{15}
%\rho=3H^2 +\frac{f}{2}  +6H^2f_T
%\end{equation}
%\begin{equation}
%\label{16}
%p=-2\dot{H}(1 + f_T -12H^2f_{TT}) - (3H^2 +\frac{f}{2}+6H^2f_T)
%\end{equation}
Using equation \eqref{12}, \eqref{13} and \eqref{14} we can write the equation of state parameter (EoS) as follows
\begin{equation}
\label{15}
\omega=\frac{p}{\rho}=-1-\frac{4\dot{H}(1 + f_T -12H^2f_{TT})}{(6H^2 +f+12H^2f_T)}
\end{equation}
where $8\pi G=1$.
Also, the matter fluid satisfies the continuity equation
\begin{equation}\label{16}
\dot{\rho}+3H(1+\omega)\rho=0,
\end{equation}
 which can be used to study the dynamics of matter fluid.\\
 
\section{Kinematic variables}\label{III}

\begin{figure}[H]
  \centering
  \includegraphics[width=8.5 cm]{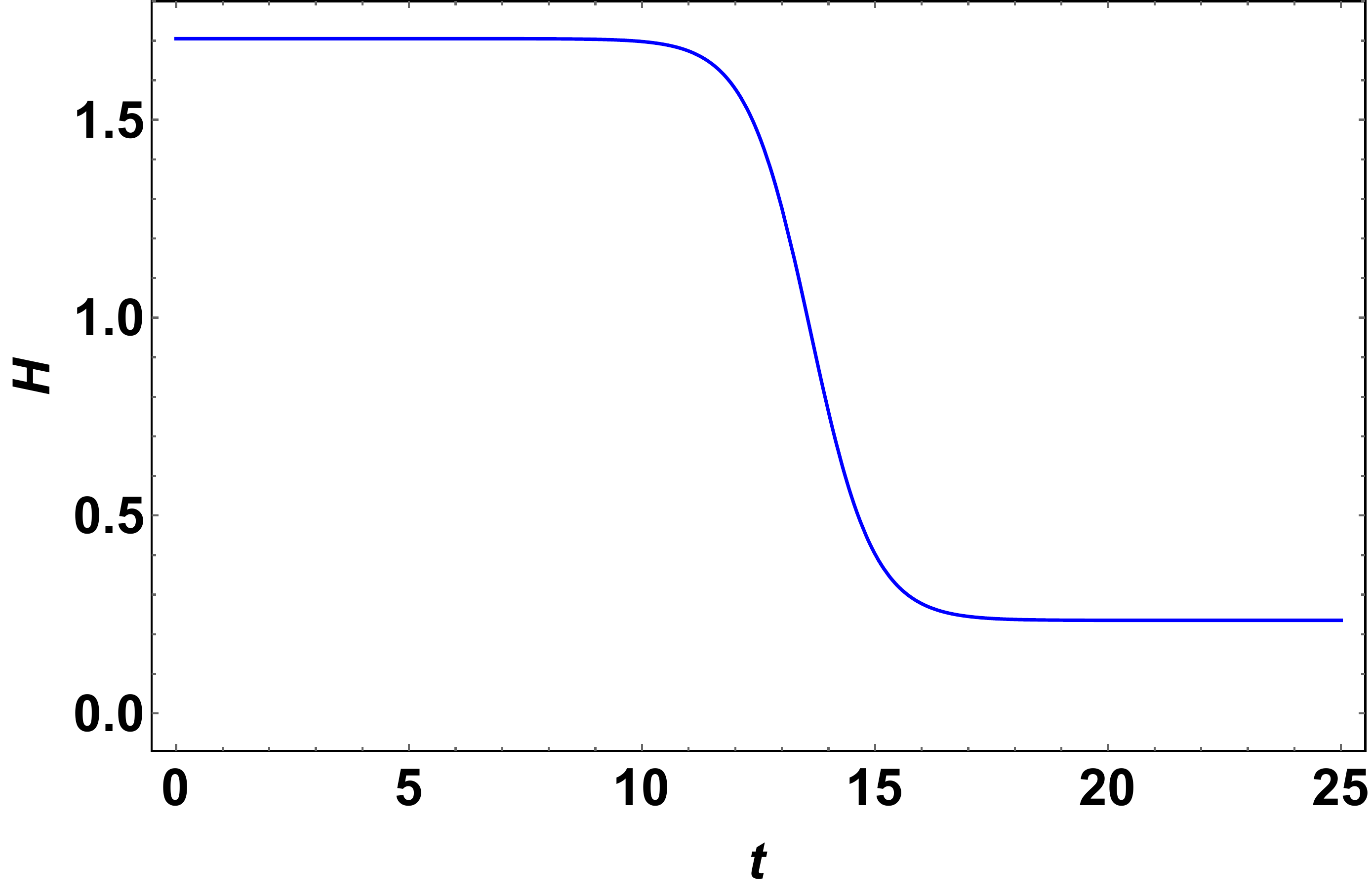}
  \caption{Plot of Hubble parameter $(H)$ as a function of cosmic time $(t)$ for $c=0.97,d=1,m=0.735,$ \& $n=10$.}
  \label{f1}
\end{figure}

\begin{figure}[H]
  \centering
  \includegraphics[width=8.5 cm]{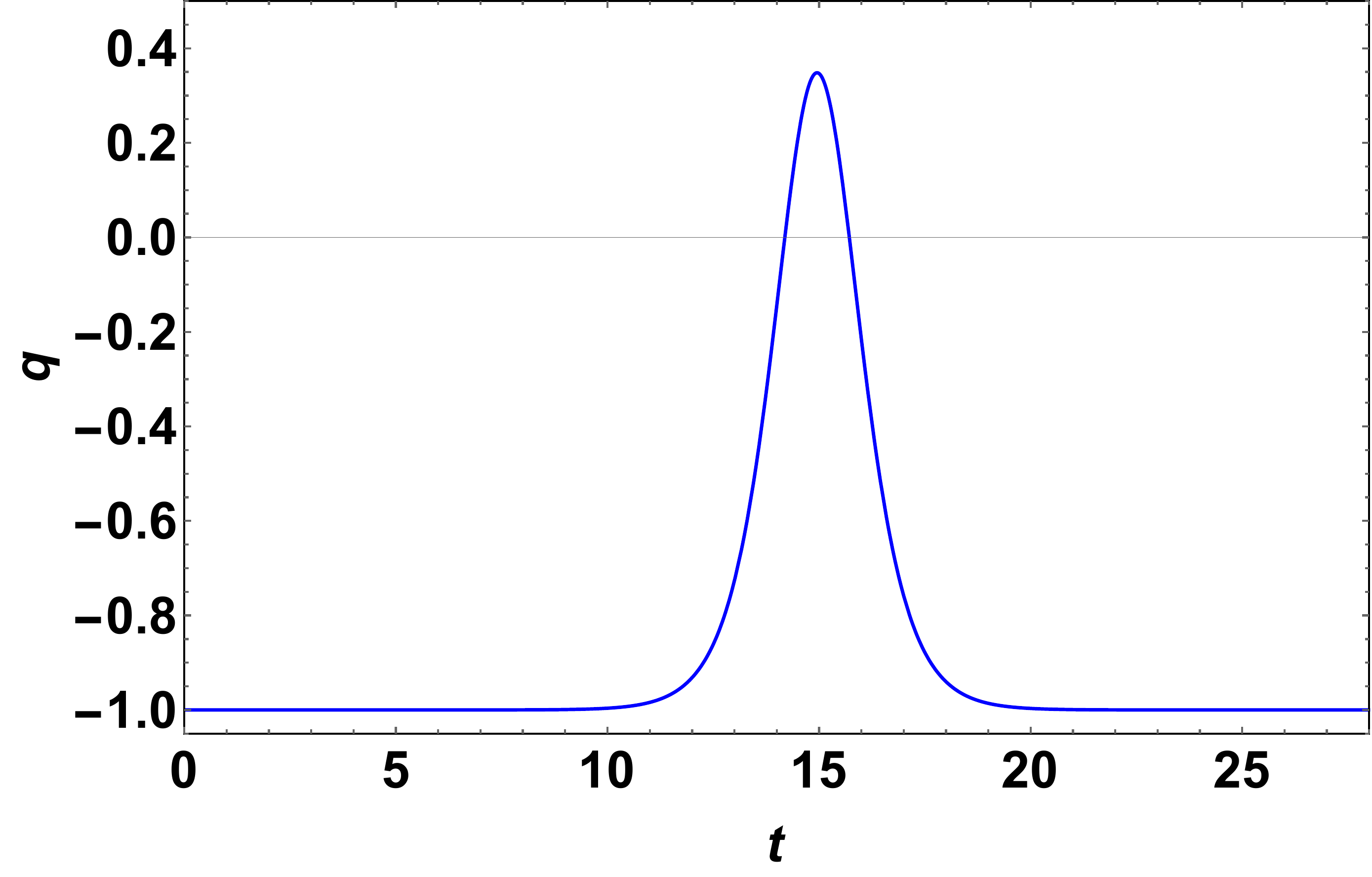}
  \caption{Plot of deceleration parameter $(q)$ as a function of cosmic time $(t)$ for $c=0.97,d=1,m=0.735,$ \& $n=10$.}
  \label{f2}
\end{figure}

The cosmological parameters such as scale factor $a(t)$, Hubble parameter $H(t)$, deceleration parameter $q(t)$ have a very significant role in describing the evolution of the Universe. And, these are the key parameters of most of the cosmological models in modified gravity theories. For analysis, we have presumed the scale factor presented by Moraes and Santos \cite{Moraes/2016a} as follows 

\begin{equation}
\label{18}
a(t)=e^{c t} [\text{sech}(n-m t)]^d.
\end{equation}
The motivation of taking this scale is that it shows a complete cosmological scenario.\\
Using \eqref{18}, one can get the Hubble parameter $H(t)$ and deceleration parameter $q(t)$ as follows
\begin{equation}
H(t)=\frac{\dot{a}}{a}= c+d m \tanh (n-m t),
\end{equation}
\begin{equation}
q=-1-\frac{\dot{H}}{H^2}=-1+\frac{d m^2}{[c \cosh (n-m t)+d m \sinh (n-m t)]^2}.
\end{equation}
Currently, our universe is undergoing an accelerated expansion phase for that the second derivative of the scale factor i.e., $\ddot{a}$ must be positive or $\dot{a}$ is an increasing function over the cosmic time evolution. Additionally, the Hubble parameter $H(t)$ is a decreasing function over the growth of time. From Fig. \ref{f1}, one can easily see that $H(t)$ keeps its value approximately the same in the early stage of the universe. After that, it gradually decreases and maintains a constant behaviour during the late time of the universe.  According to standard cosmology, the Hubble parameter is proportional to the energy density in the late time of the cosmic evolution. Luckily, we get the same behaviour of the Hubble parameter for the late time of cosmic evolution. \\
The evolution of the deceleration parameter as a function of cosmic time presented in Fig. \ref{f2}. From Fig. \ref{f2}, one can observe that the evolution of the deceleration parameter starts at $q=-1$, which represents the de-Sitter expansion phase, and then it goes to the deceleration phase through accelerating power-law expansion phase. After that, it again returns to the de-Sitter expansion phase in the late time.

\section{Cosmological model in $f(T)$ gravity}\label{IV}

\begin{figure}[H]
  \centering
  \includegraphics[width=8.5 cm]{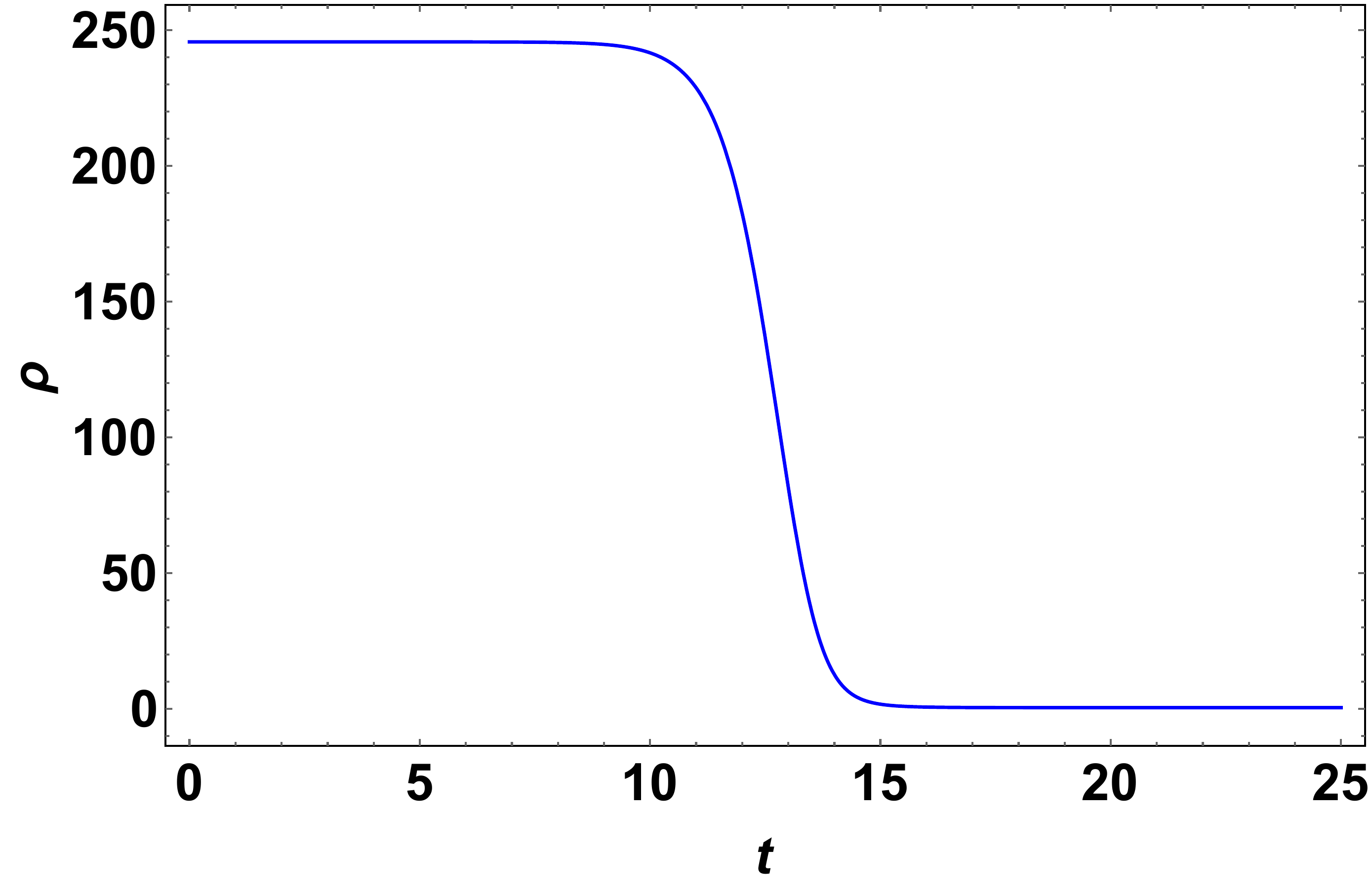}
  \caption{Plot of energy density $(\rho)$ as a function of cosmic time $(t)$ for $c=0.97,d=1,m=0.735,n=10, \alpha=2,$ \& $\beta=-0.5$.}
  \label{f3}
\end{figure}

\begin{figure}[H]
  \centering
  \includegraphics[width=8.5 cm]{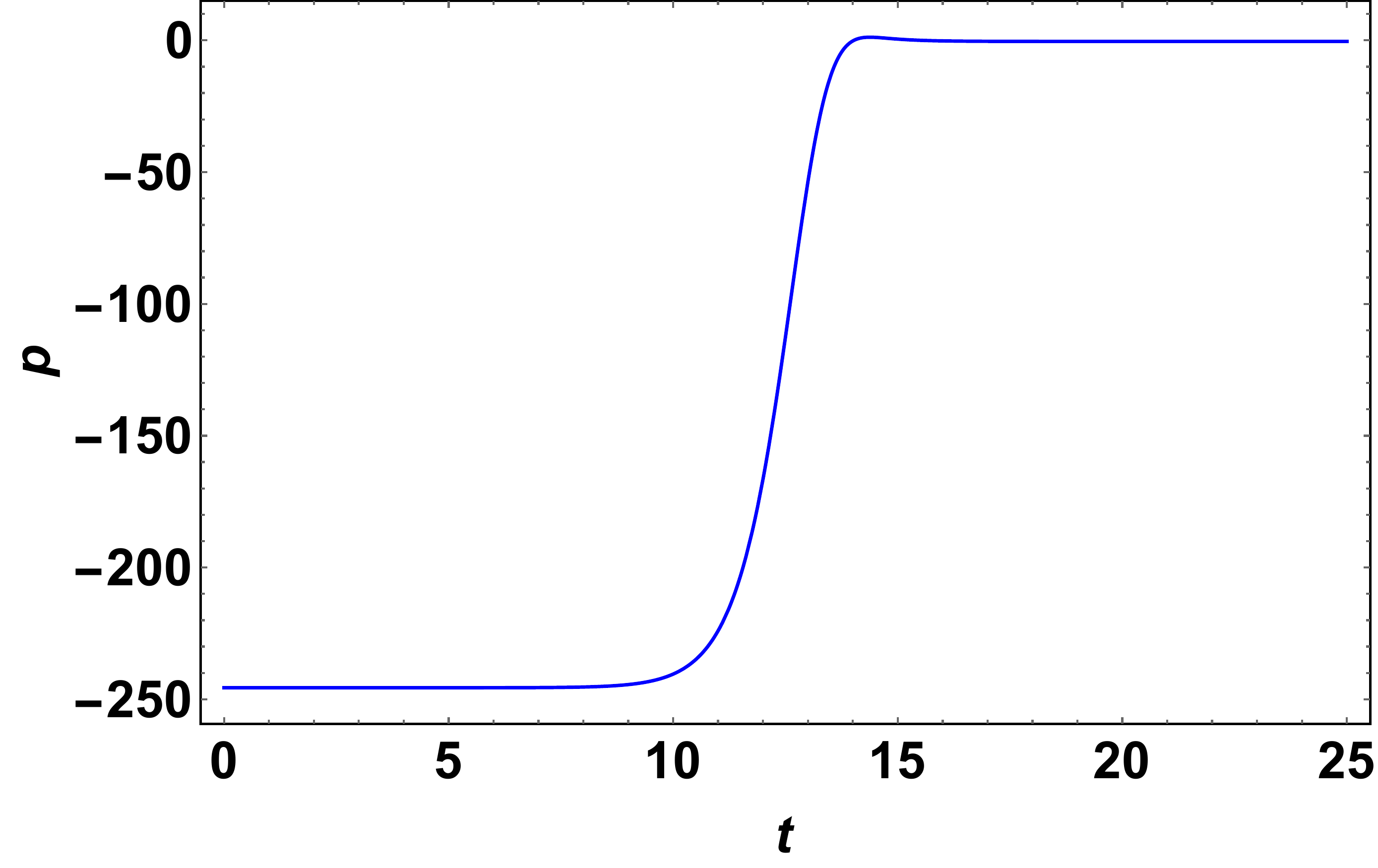}
  \caption{Plot of pressure $(p)$ as a function of cosmic time $(t)$ for $c=0.97,d=1,m=0.735,n=10, \alpha=2,$ \& $\beta=-0.5$.}
  \label{f4}
\end{figure}

\begin{figure}[H]
  \centering
  \includegraphics[width=8.5 cm]{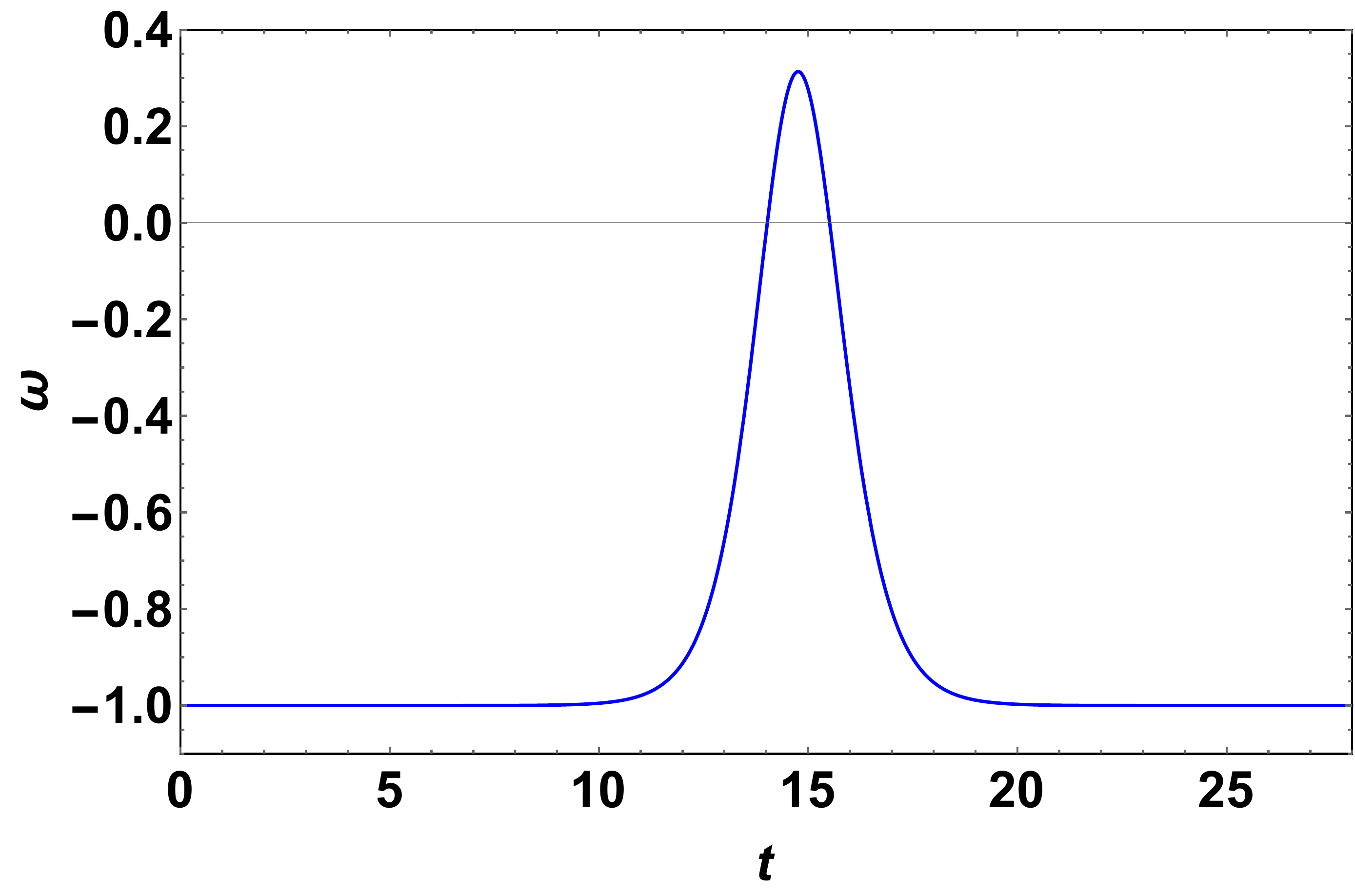}
  \caption{Plot of EoS parameter $(\rho)$ as a function of cosmic time $(t)$ for $c=0.97,d=1,m=0.735,n=10, \alpha=2,$ \& $\beta=-0.5$.}
  \label{f5}
\end{figure}

In this section, we presume the general function of $f(T)$ to analyse the cosmological model in teleparallel gravity as follows
\begin{equation}
\label{17}
f(T)=T+\beta(-T)^{\alpha}.
\end{equation}
Using \eqref{17}, \eqref{18} in \eqref{12}, \eqref{13} and \eqref{15} we can find the energy density $\rho$, pressure $p$ and EoS parameter $\omega$ as follows
\begin{widetext}
\begin{equation}
\label{19}
\rho= 6 [c+d m \tanh (n-m t)]^2-\frac{\beta}{2}6^{\alpha }( 2\alpha -1)  \left[c+d m \tanh (n-m t)\right]^{2\alpha },
\end{equation}
\begin{multline}
\label{20}
p=6^{\alpha-1 } (2 \alpha -1) \beta  [c+d m \tanh (n-m t)]^{2\alpha-2 } \left[3 c^2+d m \tanh (n-m t) \lbrace 6 c+m (2 \alpha +3 d) \tanh (n-m t)\rbrace -2 \alpha  d m^2\right]\\
-2 \left[3 c^2+d m \tanh (n-m t) \lbrace 6 c+(3 d+2) m \tanh (n-m t)\rbrace-2 d m^2\right],
\end{multline}
and
\begin{equation}
\label{21}
\omega=-1-\frac{2 d m^2 \text{sech}^2(n-m t) \left\lbrace 6^{\alpha } \alpha  (2 \alpha -1) \beta  \left[c+d m \tanh (n-m t)\right]^{2\alpha }-12 [c+d m \tanh (n-m t)]^2\right\rbrace}{36 [c+d m \tanh (n-m t)]^4-2^{\alpha } 3^{\alpha +1} (2 \alpha -1) \beta  \left[c+d m \tanh (n-m t)\right]^{2\alpha +2}}.
\end{equation}
\end{widetext}
The profile of energy density $\rho$, pressure $p$, and equation of state parameter $
\omega$ for the cosmic time shown in Fig. \ref{f3}, \ref{f4}, and \ref{f5}, respectively. From Fig. \ref{f3}, one can easily observe that the energy density is high in the early time of the universe and then gradually decreases to null. The pressure $p$ lies in the negative range to suffice the acceleration of the universe in Fig. \ref{f4}. We got an interesting behaviour of the EoS parameter $\omega$, as shown in Fig. \ref{f5}. In the early phase of the evolution of the universe $\omega$ takes its values $-1$ and smoothly raises to its maximum value $\omega\sim \frac{1}{3}$, and finally converging to $-1$ in order to serve the recent observation \cite{Riess/1998,Perlmutter/1999,deBernardis/2000,Colless/2001,Perlmutter/2003,Spergel/2003,Peiris/2003,Tegmark/2004,Cole/2005,Springel/2006,Astier/2006,Riess/2007,Spergel/2007,Ade/2014,Komatsu}. From Fig. \ref{f5}, we concluded that the universe starts with the acceleration smoothly, goes to the deceleration phase, and finally returns to its second phase of accelerated expansion.

\section{Statefinder Diagnostics}\label{sd}

\begin{figure}[H]
  \centering
  \includegraphics[width=8.5 cm]{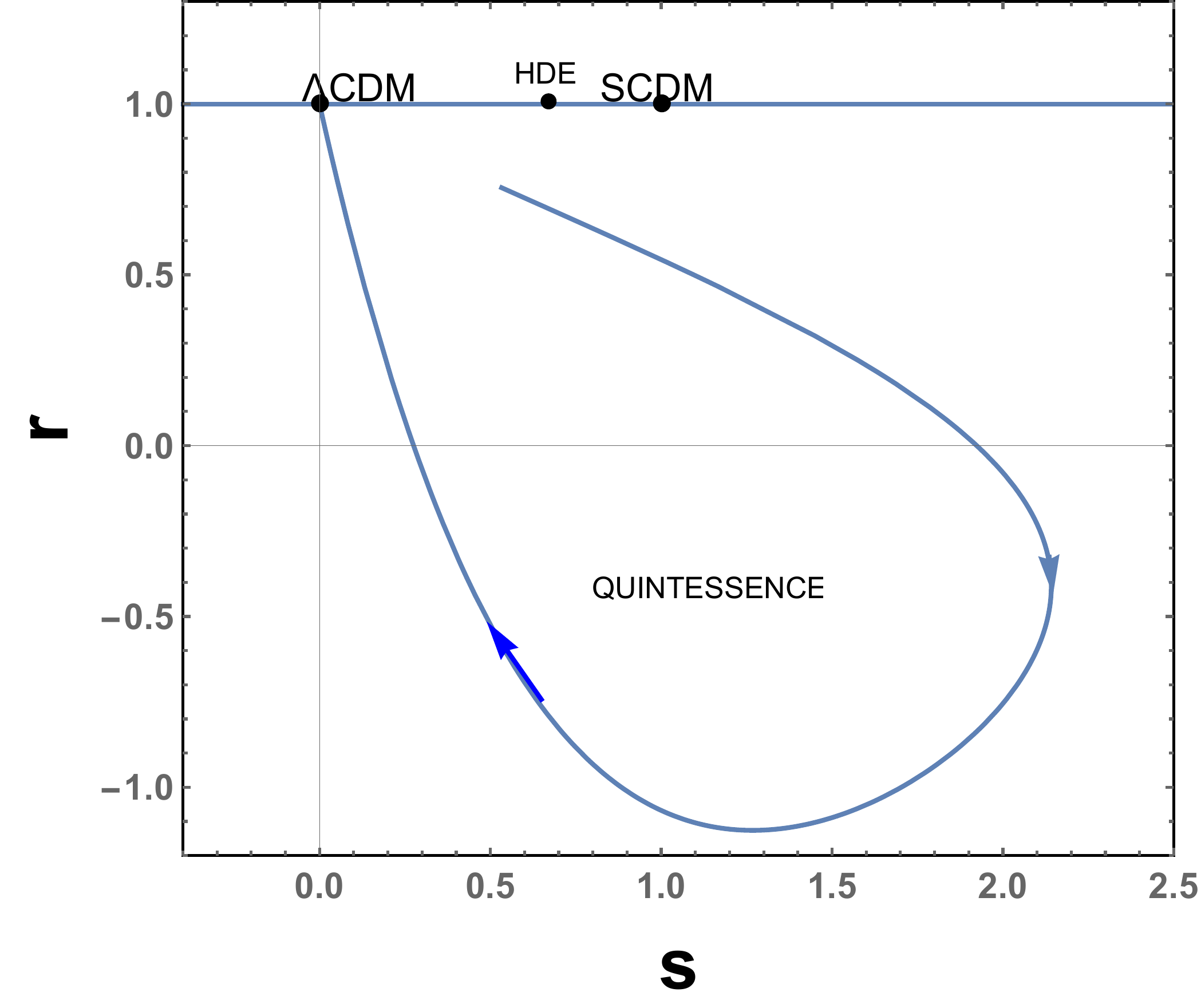}
  \caption{$r-s$ parametric plot for $c=0.97,d=1,m=0.735,$ \& $n=10$.}
  \label{f6}
\end{figure}

\begin{figure}[H]
  \centering
  \includegraphics[width=8.5 cm]{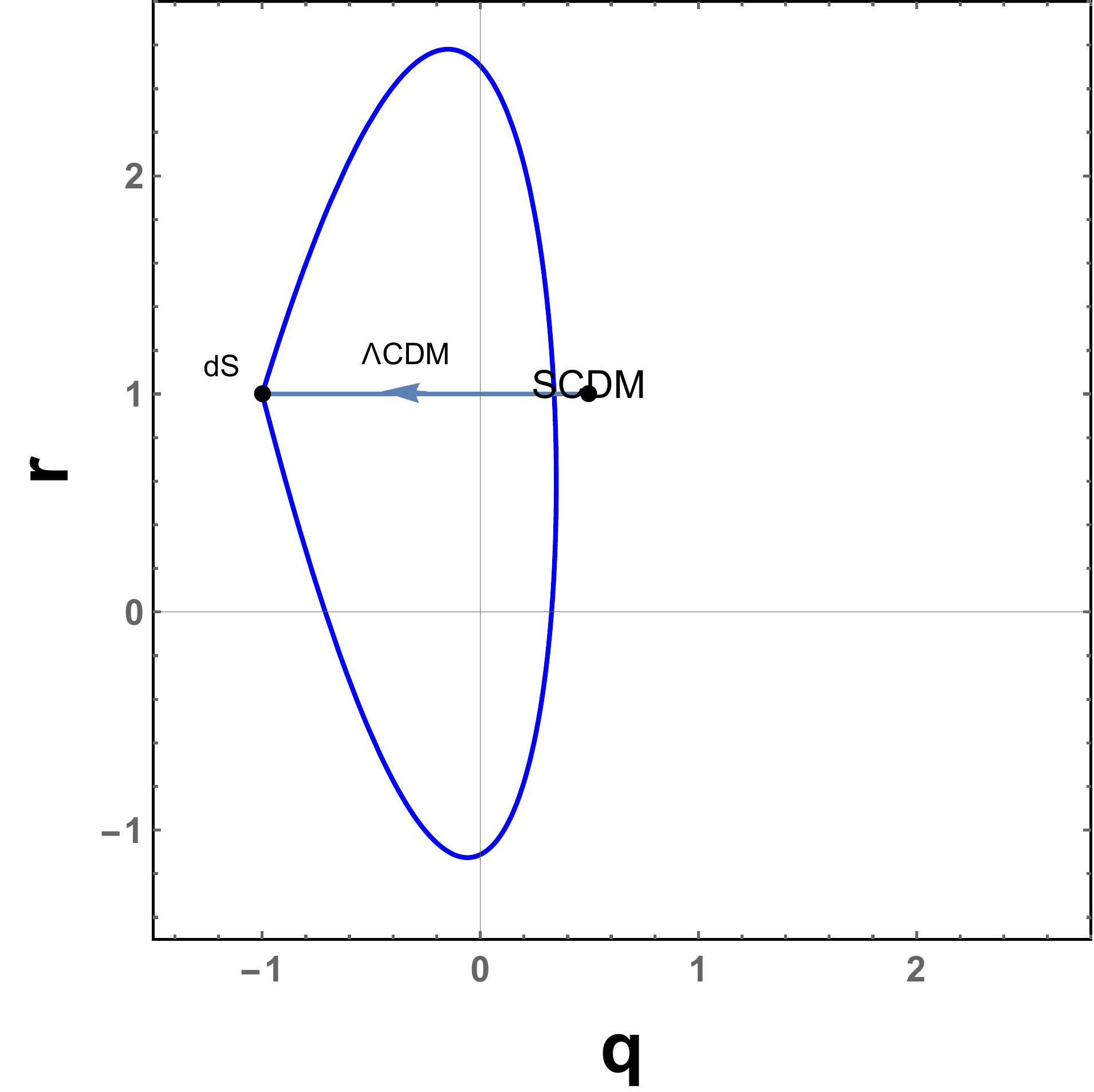}
  \caption{$r-q$ parametric plot for $c=0.97,d=1,m=0.735,$ \& $n=10$.}
  \label{f7}
\end{figure}

The unknown nature of dark energy (DE) arises many problems in modern cosmology. To understand the nature of dark energy, many dark energy models such as $\Lambda$CDM, HDE, SCDM, CG, quintessence have proposed in the literature. Also, these DE models have different behaviours in comparison to each other. Therefore, the $\left\lbrace r,s\right\rbrace$ parametrization technique proposed in \cite{Sahni/2003,Alam/2003} is also used to distinguish all these DE models. The $r$ and $s$ take the value as follows
\begin{equation}
\label{24}
r=\frac{\dot{\ddot{a}}}{aH^3},
\end{equation}
\begin{equation}
\label{25}
s=\frac{2}{3}\frac{r-1}{2q-1},
\end{equation}
where $q\neq \frac{1}{2}$.
The pair $\left\lbrace r,s\right\rbrace$ represents different dark energy models. These are discussed in following
\begin{itemize}
\item The pair $\left\lbrace r=1,s=0\right\rbrace\rightarrow$ $\Lambda$CDM .

\item The pair $\left\lbrace r=1,s=1\right\rbrace\rightarrow$ SCDM.

\item The pair $\left\lbrace r=1,s=\frac{2}{3}\right\rbrace\rightarrow$ HDE.

\item The pair $\left\lbrace r>1,s<0\right\rbrace\rightarrow$ CG.

\item The pair $\left\lbrace r<1,s>0\right\rbrace\rightarrow$ Quintessence.
\end{itemize}
The main idea is to study the convergence and divergence nature of the trajectory of the $r-s$ parametric curve corresponding to the $\Lambda$CDM model. The deviation from $\left\lbrace 1,0 \right\rbrace$ represents the deviation from $\Lambda$CDM model. Furthermore, the values of $r$ and $s$ could be concluded from the observation \cite{Albert,Albert/2}. Therefore, it is worthy of describing the DE models in the near future.\\
Using \eqref{18} in \eqref{24}, \eqref{25} we can rewrite the $r,s$ as follows 
\begin{widetext}
\begin{equation}
\label{26}
r=\frac{c^3+d m \tanh (n-m t) \left\lbrace c^2+(d+1) m \tanh (n-m t) [3 c+(d+2) m \tanh (n-m t)]-(3 d+2) m^2\right\rbrace-3 c d m^2}{[c+d m \tanh (n-m t)]^3},
\end{equation}
\begin{equation}
\label{27}
s=\frac{2 d m^2 \text{sech}^2(n-m t) [3 c+(3 d+2) m \tanh (n-m t)]}{3 [c+d m \tanh (n-m t)] \left\lbrace c^2+d m \tanh (n-m t) [6 c+(3 d+2) m \tanh (n-m t)]-2 d m^2\right\rbrace}.
\end{equation}
\end{widetext}
In Fig. \ref{f6}, the parametrization of $r$ and $s$ shown in $(r,s)$ plane and the arrow mark represents the direction of the trajectory. From Fig. \ref{f6}, one can observe that the trajectory diverges from $\Lambda$CDM model, initially and later, it converges to $\Lambda$CDM model. Also, the evolution of the trajectory completely lies in the quintessence. Additionally, we have shown the parametrization of $r$ and $q$ in Fig. \ref{f7}. From Fig. \ref{f7}, we observed that our model starts with the de-Sitter universe, and initially, it goes to Chaplygin gas (which is represented by $r>1$). After that, it comes to quintessence and finally converges to the de-Sitter universe.

\section{Energy conditions}\label{ecs}

There are many more advantages to study the energy conditions such as to understand singularity theorem of space-like and time-like curves, to describe geodesics, black holes, and wormholes based on the well known Raychaudhuri equation \cite{Moraes/2017a,Wald/1984}. Energy conditions in teleparallel gravity have been studied in \cite{Liu/2012,Azizi/2017,Zubair/2015,Mandal/2020}. Also, in the presence of singularity, energy conditions provide us the edge of the parameters. The energy conditions are the linear relations between energy density and pressure. They are presented follows
\begin{itemize}
\item SEC : $\rho+3p\geq 0$;

\item WEC : $\rho\geq 0, \rho+p\geq 0$;

\item NEC : $\rho+p\geq 0$;

\item DCE : $\rho\geq 0,|p|\leq \rho$.
\end{itemize}
\begin{figure}[H]
  \centering
  \includegraphics[width=8.5 cm]{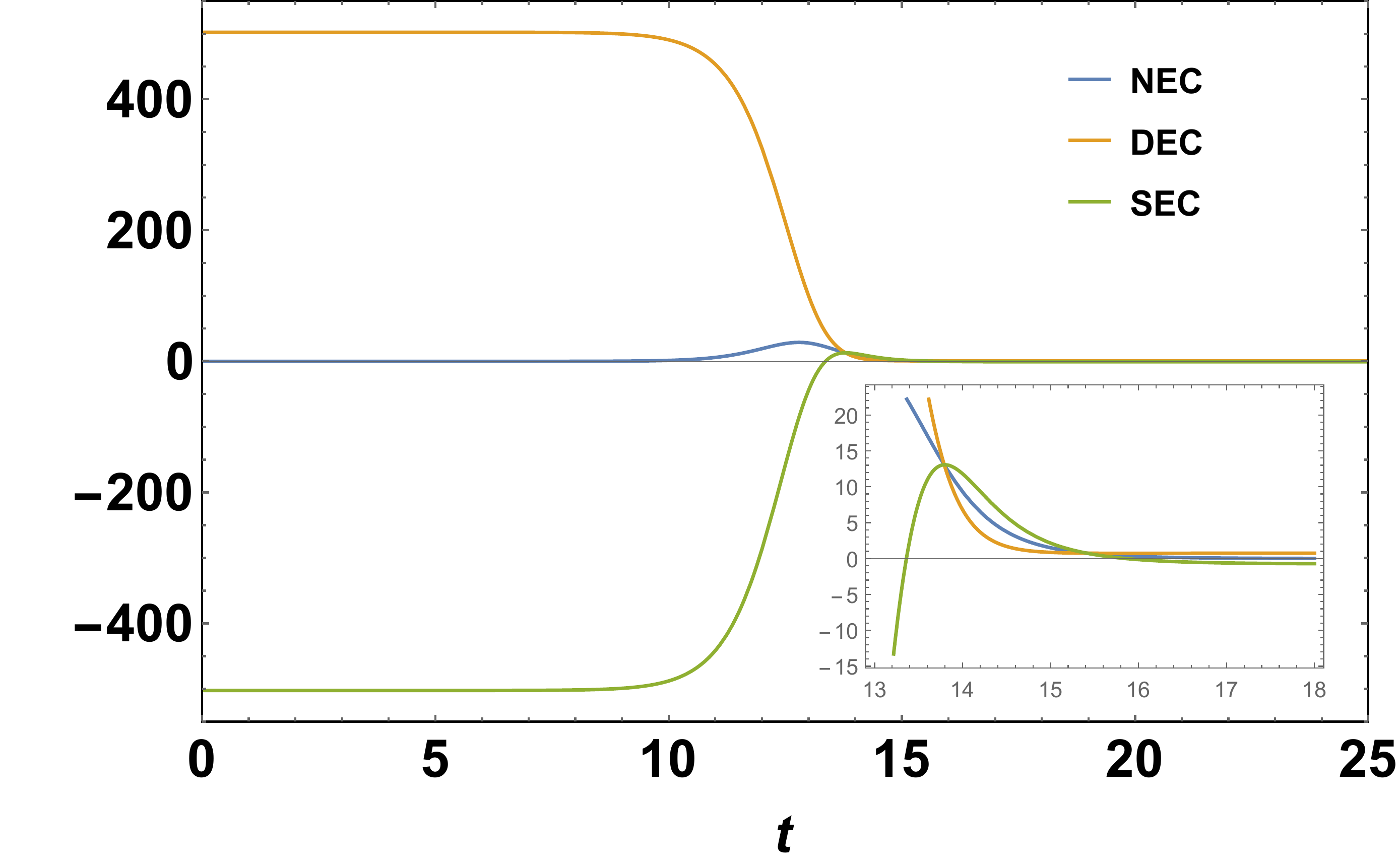}
  \caption{ECs as a function of $t$ for $c=0.97,d=1,m=0.735,n=10, \alpha=2,$ \& $\beta=-0.5$.}
  \label{f8}
\end{figure}
In the above figure, the portrait of all Energy conditions (ECs) shown. From Fig. \ref{f8}, we observed that the WEC, NEC satisfies in our present model, whereas SEC violates. In detail, if one can see close to the behaviour of SEC, then he will observe for a time interval, it satisfies by our model. As we discussed the profile of $\omega$ previously, it starts with acceleration, then smoothly goes to the deceleration phase and finally returns to the accelerated phase. The exciting thing is getting the same results for SEC.

\section{Results and conclusions}\label{f}

The article represents a complete cosmological scenario of the FLRW universe in teleparallel gravity. To find the exact solution of the field equations, we considered the general function $f(T)=T+\beta(- T)^{\alpha}$ which is proposed in \cite{Barrow/2016}. In the below paragraphs, we shall discuss the cosmological feasibility of the fundamental of $f(T)$ gravity.\newline
The EoS parameter $\omega$ is a key parameter to describe the different matter-dominated eras in the evolution of the universe. The portrait of the EoS parameter with time $t$ have shown in Fig. \ref{f5}. By analyzing Fig. \ref{f5}, we observed that in the early phase of the evolution of the universe, the EoS parameter takes its value as $\omega\sim -1$. This result is in harmony with some limitations of EoS, which have put to the inflationary EoS, recently \cite{Bamba/2013,Ackerman/2011,Aldrovani/2008}. After inflation, $\omega$ smoothly rises to $\sim \frac{1}{3}$, and this is the maximum value of $\omega$ during its evolution. Moreover, $\omega\sim \frac{1}{3}$ represents the radiation-dominated phase of the evolution of the universe \cite{Ryden/2003,Dodelson/2003}. Also, we noted that the EoS parameter $\omega$ converges to $-1$ in the late time. And, $\omega=-1$ is the present value of the universe for the recent cosmic acceleration which is observed by the anisotropies in the temperature of cosmic microwave background radiation \cite{Hinshaw/2013,Planck,Haridasu/2018,Garza/2019,Lin/2019,Jesus}.\newline
The overall picture of the EoS parameter shows that the universe started from the inflation with $\omega=-1$ (as the negative values of $\omega$ present the acceleration of the universe), then smoothly it takes its value to positive i.e., deceleration phase. Finally, it returns to the accelerated phase which is the second phase of the acceleration of the universe in the present time.\newline
In section \ref{ecs}, we show the temporal evolution of the energy conditions (ECs). It is to keep in mind that to serve the late-time acceleration of the universe, the SEC has to violate \cite{Planck}. As we discussed in the previous section for the late time acceleration $\omega$ takes its value as $-1$, so the SEC $\rho(1+3\omega)<0$ always. From Fig. \ref{f8}, one can quickly observed that NEC, WEC satisfies, whereas SEC violates. Also, we noted that initially SEC violates, then it satisfies for a time interval and again violates. This result support the result we got for $\omega$.\newline
In Fig. \ref{f3}, \ref{f4}, we show the behaviour of energy density and pressure for our model. The energy density should be positive to fulfil WEC, and pressure should be negative for cosmic acceleration. The evolution of pressure and energy density of our model satisfies WEC. And, this type of property only achieved by the modification of general relativity or exotic matter.\newline
In section \ref{sd}, we investigate the difference between the dark energy models with our model by statefinder diagnostic. From Fig. \ref{f6}, we observed that the trajectory of $\left\lbrace r,s\right\rbrace$ parametric curve deviates from $\Lambda$CDM model initially. However, at the late time, it coincides with the $\Lambda$CDM model, which is consistent with standard cosmology. Addition to this, we show $\left\lbrace r,q\right\rbrace$ parametric plot in Fig. \ref{f7}. It is also consistent with the standard cosmology. 

\section*{Acknowledgments}  S.M. acknowledges Department of Science \& Technology (DST), Govt. of India, New Delhi, for awarding Senior Research Fellowship (File No. DST/INSPIRE Fellowship/2018/IF180676). We are very much grateful to the honorable referee and the editor for the illuminating suggestions that have significantly improved our work in terms of research quality and presentation.

\end{document}